\documentclass[seceq]{ptptex}
\usepackage{wrapft}

\usepackage{graphicx}

\pubinfo{Vol.~118, No.~2, 2007}

\title{
Modeling Dense Stellar Systems
}

\author{
Piet \textsc{Hut}$^{1,}$\footnote{E-mail: piet@ias.edu},
Shin
  \textsc{Mineshige}$^{2,}$\footnote{E-mail: minesige@yukawa.kyoto-u.ac.jp},
Douglas C. \textsc{Heggie}$^{3,}$\footnote{E-mail: d.c.heggie@ed.ac.uk},
and Junichiro \textsc{Makino}$^{4,}$\footnote{E-mail: makino@yso.mtk.nao.ac.jp}
}

\inst{
$^1$Institute for Advanced Study, 1 Einstein Drive, Princeton, NJ 08540, USA

$^2$Yukawa Institute for Theoretical Physics, Kyoto University,
      Kyoto 606, Japan

$^3$School of Mathematics and Maxwell Institute for Mathematical Sciences,
      University of Edinburgh, King's Buildings, Edinburgh EH9 3JZ,
      Scotland, U.K.

$^4$National Astronomical Observatory of Japan, Osawa 2-21-1, Mitaka,
Tokyo 181-8588, Japan
}

\recdate{
x/x/2007}

\abst{
Black holes and neutron stars present extreme forms of matter that
cannot be created as such in a laboratory on Earth.  Instead, we have
to observe and analyze the experiments that are ongoing in the
Universe.  The most telling observations of black holes and neutron
stars come from dense stellar systems, where stars are crowded close
enough to each other to undergo frequent interactions.  It is the
interplay between black holes, neutron stars and other objects in a
dense environment that allows us to use observations to draw firm
conclusions about the properties of these extreme forms of matter,
through comparisons with simulations.  The art of modeling dense
stellar systems through computer simulations forms the main topic
of this review.
}

\begin{document}

\maketitle

\section{Invitation}

If you are a student looking for a thesis topic, or a researcher
looking for a new field to explore, you may be interested to consider
a relatively new area of study, dense stellar systems.  Defined as
regions where stars are so close together that they frequently collide,
dense stellar systems give rise to all kinds of interesting phenomena.
Here are some reasons to consider dense stellar systems:

\begin{itemize}
\item{if you are interested in fundamental physics, and in extreme
   forms of matter such as neutron stars and black holes; and if you
   would like to know where to find them, how to observe them, and how
   to interpret those observations.}
\item{if you are interested in making fundamental contributions to
   astrophysics, and you are afraid that all basic discoveries have
   been made already, then consider dense stellar systems, as an
   interdisciplinary field where many of the basic questions have not
   even been addressed, let alone answered.}
\item{if you share Newton's interest in the classical N-body problem,
   something he didn't have the computational tools for, dense stellar
   systems offer you the closest application in the real world for
   this abstract problem, fascinating in its simplicity of
   formulation, as well as its complexity of behavior.}
\end{itemize}

\noindent
As an example of the second point, we can take stellar evolution.  The
most basic stellar evolution calculations using electronic computers
were first performed in the 1950s, and then developed in great detail
in the 1960s.  The next four decades have mainly seen refinements of
modeling techniques, including the difficult treatment of mass transfer
in binary systems, but the text books of the 1960s still form a good
entry point for learning about the basic approach to the evolution of
isolated single stars and normal binary systems.

In contrast, a full stellar evolution modeling of a dense stellar
system has yet to be carried out.  The very first attempt in that
direction was made recently in a PhD thesis by Ross Church in
2006\cite{rf:1}.  With respect to evolving the ecological system of an
interacting star cluster, we are in a similar state as where Martin
Schwarzschild was in the early 1950s, when he first started to use
John von Neumann's computer in Princeton to follow the evolution of a
single star\cite{rf:2}.  In short, the exploration of the new frontier
of dense stellar systems has just begun and is inviting you to join
in the adventure.

The structure of this review is as follows.  Section 2 offers five
different perspectives on dense stellar systems, from the point of
view of fundamental physics, astrophysics, classical physics,
computational physics, and interdisciplinary physics collaborations.
Section 3 describes how neutron stars and black holes can be detected
through radiation emitted from their vicinity.  Section 4 introduces
different places in the Universe where dense stellar systems can be
found, such as globular star clusters and galactic nuclei, with an
emphasis on our own galactic center.  Section 5 highlights the
multi-scale and multi-physics challenges that simulations of dense
stellar systems face, and also mentions the use of GRAPE special
purpose hardware.

\section{Perspectives on Dense Stellar Systems}

Right at the center of our galaxy, a massive black hole resides,
surrounded by a dense cluster of stars.  To get a sense of how crowded
the central region is, let us consider the contents of the inner
parsec (pc) of our galaxy, an area that is also called the central
nucleus.  For comparison, in the neighborhood of our Sun, the distance
between individual stars is typically more than 1 pc
($\approx3\times10^{16}$m).
In contrast, a sphere with a radius of 1 pc around the center of our
galaxy contains a black hole with a mass of more than three million
solar masses, together with a similar amount of mass in stars that
move in tight orbits around the central black hole.

The density of stars in the center of our galaxy is thus more than a
million times higher than that in the neighborhood of our Sun.  And
there are other places as well, that have a much higher density then
our local neighborhood.  The centers of some globular clusters, too,
approach a similar density.  In such cases, it is unavoidable that
many stars undergo close encounters and even physical collisions, with
high probability, within their life times.  It is these environments,
called dense stellar systems, that we will focus on in this review.
To start with, we will look at these systems from a variety of
different perspectives.

Our Sun has not always been as isolated as it is now.  Most likely,
it was born in a much denser `nest' of stars.  Recently, direct
evidence for the formation of the Sun in a dense stellar system has
been obtained by isotope analysis of meteorites in the solar system,
hinting at the presence of at least one supernova very close to the
young Sun\cite{rf:2a}.

\subsection{Fundamental Physics Perspective}

Black holes constitute the most extreme form of matter known in the
universe.  According to the best tested theory of gravity that we
have, general relativity, the matter in a black hole is compressed in
a central singularity, at an infinitely high density.  Most likely,
this description is only an approximation: whenever a theory in
physics predicts the occurrence of singularities, it is a sign that
the theory itself breaks down, and has to be replaced by a more
detailed description, more appropriate for the area under concern.

Although there are many speculations as to the type of quantum gravity
theory that might replace general relativity, we do not yet know which
theory is the correct one.  Therefore, astrophysicists continue to rely
on general relativity as their best guide.  There is an additional
reason to do so: the disagreements between a quantum gravity theory
and general relativity are likely to be confined to an area very close
to where general relativity predicts a central singularity.  This area
lies deep within the event horizon, the area from which no light and
no other classical form of information can escape.

As a consequence, effects that are observationally accessible have to
occur outside the event horizon, where the classical approximation is
expected to be highly accurate.  Even so, observational tests are very
important in this regime.  So far, general relativity has been tested
largely in the weak-field approximation.  Observations of phenomena
just outside the horizon are still mostly unexplored, and would form a
very welcome addition to test our most basic theory of gravitation.

Neutron stars are another example of extreme objects.  Unlike black
holes, they are made out of conventional matter, but in a very extreme
form.  The density of a neutron star is comparable to the density of
an atomic nucleus.  Since the diameter of a nucleus is about $10^{-5}$
of the diameter of an atom, the density of such material is roughly
$10^{15}$ times larger than that of water.  If the mass of the Sun
would be compressed to form a neutron star, its diameter would be
about 10 km, only a few times larger than the Schwarzschild radius of
the Sun which is roughly 3 km.

Both black holes and neutron stars can generate copious gravitational
waves when they collide and merge.  In order to predict the frequency
and characteristics of such merger events, simulations of dense
clusters of stars systems play an essential role in the ongoing
efforts at detections of gravitational waves.  In general, all
these phenomena, from nuclear matter in bulk to event horizons
and gravitational waves, cannot be created in a laboratory on Earth.
Therefore, we have to make do with the laboratories that nature
provides us, in the form of dense stellar systems, which leads us
to a switch from fundamental physics to astrophysics.

\subsection{Astrophysics Perspective}

When massive stars undergo a supernova explosion at the end of their
life, they may produce a black hole or a neutron star as a remnant.
Such a remnant is difficult to observe, a black hole because it is
black, a neutron star because its size is so small that the thermal
radiation emitted well after its birth is hard to detect.  Depending
on its magnetic field and spin rate, a neutron star can be visible at
radio wavelengths as a pulsar, but in that case, too, gradual spindown
will let the pulsar become invisible on a time scale that is short
compared to the age of the galaxy.

There are two ways to make those extreme objects visible.  There is an
interesting parallel here with particle physics, where exotic particles
are also studied using two ways, either by letting them collide with
each other, or by studying their bound states.  In astrophysics,
collisions between stars can make otherwise invisible objects light up.
In addition, a binary star containing one normal star and one extreme
object can produce bright X rays when matter from the normal star falls
onto the compact object.

Most stars in the Universe never interact very strongly with other
stars, at least during their adult life, after they have left the
interstellar gas cloud that was the nest in which they were born.
However, there are various `dense stellar systems', such as globular
clusters and galactic nuclei, that contain stars that are sufficiently
close to their neighbors to make collisions quite likely.  By modeling
the structure of such dense stellar systems, and comparing the modeling
results with observations, we can gain valuable information about the
nature of extreme objects, such as neutron stars and black holes.  For
example, only by studying the full ecology of a dense star system can
we interpret the properties of the bound states between stars, in the
form of binaries that may contain compact objects.

In addition to such stellar-mass objects, formed as byproducts of
stellar evolution, many galaxies contain far more massive black holes
at their centers.  Our own galaxy contains a central black hole with a
total mass close to $3.6 10^6$ solar masses.  This is a rather modest
central black hole.  Some galaxies contain holes that are more than a
billion times more massive than our Sun.  Such a central black hole
becomes detectable only through interactions with the environment.
Gas that is lost from nearby stars, or even stars plunging into such a
supermassive black hole, can produce radiation in the X ray range as
well as other wave length bands.  In addition, a sufficiently massive
black hole also affects the distribution and kinematics of the stars
around it.  Either way, the study of dense stellar systems is
important for interpreting the observations of galactic nuclei.

\subsection{Classical Physics Perspective}

The gravitational N-body problem is the oldest unsolved problem in
physics.  After Newton formulated classical mechanics, and solved the
two-body problem, attempts to solve the three-body problem did not
lead to any practical form of a general solution.  Progress in
exploring the properties of the N-body problem had to wait till
computers were available to do the very intensive number crunching
required.

For the general gravitational N-body problem, we still cannot follow
the complete evolution beyond values of N around $1-2\times10^5$.  In
order to solve the million-body problem, we will have to wait till the
end of the next decade\cite{rf:3}.  The maximum number of particles
that we have been able to simulate in full glory started as $N=10$
around 1960, and has been growing roughly according to Moore's law,
when taking into consideration that the costs of a full-fledged N-body
simulation has a scaling that is somewhat worse than $\propto N^3$.
As a result, progress from $N=10$ to $N=10^6$ implies an increase in
computational requirements of much more than a factor $10^{15}$,
corresponding to the change from kiloflops in 1960 to Exaflops, ten
or more years from now.

These small $N$ values may come as surprising news, given the many
reports in the literature of N-body calculations with billions of
particles in the case of cosmological simulations, and tens or
hundreds of millions of particles in the case of galactic dynamics.
The reason is that the latter two types of calculations are special,
in that they use a softened approximation for the gravitational
interactions, ignoring the singular nature of close encounters, and
that cosmological simulations do not span many time steps (the
Universe is dynamically young).

In the case of the general gravitational $N$-body problem, we start
with an arbitrary configuration of $N$ stars, which equilibrate in a
few crossing times.  After having reached dynamical equilibrium,
further evolution takes place on a thermal time scale, through heat
exchange through two-body relaxation effects, as in the molecular
dynamics of the atoms in a gas.  Since the effects of two-body
encounters diminish with respect to the effects of the background
potential of a star cluster as a whole, the $N^2$ computational load
of following the interactions during one dynamical time scale is
multiplied by another factor of $N$ to form the scaling of roughly
$N^3$, alluded to above.

\subsection{Computational Physics Perspective}

The challenge of simulating a dense star cluster with a million stars
is formidable, because of the enormous ranges in spatial and temporal
scales that have to be modeled simultaneously.  The size of a globular
star cluster is measured in tens of parsecs, while the diameter of
a neutron star is measured in kilometers, a discrepancy in distance
scales of a factor $10^{15}$.  The time scale problems are even worse.
The duration of a close passage of two neutron stars is measured in
fractions of milliseconds, while the evolution of a star cluster can be
comparable to the current age of the Universe, more than ten billions
years, resulting in a discrepancy of time scales of a factor $10^{21}$.

In order to make it possible to simulate a star cluster for ten
billion years, it is necessary to introduce algorithms based on
individual time steps, an approach pioneered and developed in
great detail by Aarseth\cite{rf:4}.  An analysis of the scaling
of the computational cost of the general $N$-body problem was
provided by Hut, Makino \& McMillan\cite{rf:5}, who showed that
for $N=10^5$, direct $N^2$ methods are preferred.  In order to
reach $N=10^6$, various algorithms can be employed to make a switch
from $N^2$ to $N\log N$ scaling of inter-particle interactions, using
tree methods such as introduced by Hut \& Barnes\cite{rf:6}.

Introductory material, as well as some new ideas about using a
four-dimensional space-time perspective, can be found on the website
of the {\sl Art of Computational Science}\cite{rf:7}.  There, a switch
in perspective is presented from a notion of $N$ bodies interacting in
space to a collection of $N$ world lines in spacetime, the configuration
of which can be computed in a partially asynchronous way, using not
only individual time steps, but even individual algorithms.

In addition to the many algorithmic developments, significant speed
has been gained by the construction of special-purpose hardware, in
the form of the GRAPE family\cite{rf:8}.  With a cost-performance
ratio that is one or two orders of magnitude better than that of
commercial supercomputers, the GRAPEs have dominated simulations of
dense stellar systems for the last decade.

\subsection{Interdisciplinary Physics Perspective}

A detailed study of dense stellar systems requires the collaboration
of astrophysicists with widely different specializations and backgrounds.
Besides the multi-scale challenges summarized above, there is the
multi-physics challenge of simultaneously modeling the physical
evolution of individual stars, the hydrodynamical interactions between
neighboring stars, and the gravitational interactions of the star
cluster as a whole.

Whenever two or more stars approach each other closely, they can no
longer be treated as point masses.  Hydrodynamical calculations have
to be employed to study the deformations and exchange of energy and
angular momentum, and perhaps mass transfer or even a complete merger
between the stars.  Following those dynamical events, on a time scale
of hours and days, the stars have to be followed for far longer time
scales, of order of thousands if not millions of years, to follow the
restoration of internal thermal equilibrium;

None of these treatments form part of the standard tool set of stellar
dynamics, stellar evolution, or stellar hydrodynamics.  New ideas need
to be developed, together with new techniques and new implementations.
It is in this area that there is plenty of room for basic breakthroughs,
as mentioned in the first section of this paper.  The MODEST initiative,
for MOdeling DEnse STellar systems\cite{rf:9}\cite{rf:10}, has been
organizing dozens of workshops to guide these developments, since its
inception in 2002.

\section{Emission from Compact Objects}

In the Universe there exist extreme objects that we cannot study in
our laboratories: black holes and neutron stars.
Although both have masses comparable to or moderately larger than
the mass of the Sun, their {\lq}sizes{\rq} are extremely small;
only on the order of 10 km or so.  A big distinction between them
is that a neutron star has a solid surface, while a black hole has not.
Neutron stars can support themselves by degeneracy pressure of neutrons
against self-gravity, whereas black holes are collapsed objects
because no counteracting force is strong enough to counter their gravity.

Why are we so much interested in such compact objects?
There are several answers possible to this question, but 
a primary reason would be that
we can get information as to the extreme physics, 
physics of extremely high density and high temperature material, 
sometimes with extremely strong magnetic fields, 
in extremely large gravitational fields, 
through the study of the compact objects.
The existence of such extreme objects makes
our view of the Universe remarkably rich.  we can find these objects
in dense stellar systems, and in turn
we can also use these compact objects as a probe to study the extreme
conditions in dense stellar systems.

Then, how to detect compact objects?
The most efficient way to identify 
neutron stars or black holes is to detect X-ray emission.  
This is because 
accretion onto a region with a dimension of $r_* \sim 10^6$cm
will emit strong X-rays.
If the typical luminosity of $L \sim 10^{37-38}$ erg s$^{-1}$
is emitted as blackbody radiation from the area of $4\pi r_*^2$,
the blackbody temperature will be 
$[L/(4\pi r_*^2\sigma)]^{1/4}\sim 10^7 {\rm K}$,
(where $\sigma$ is the Stefan-Boltzmann constant)
which implies the emission of X-rays.
(If neutron stars have strong magnetic fields
radio emission is also very important, in addition to X-ray emission.)
If some source emits intense X-ray (and radio) emission and 
if the emission region is compact, 
it is likely to be an X-ray binary system, where a normal star loses
gas that accretes onto the compact object.
Let us see, next, some more details of emission properties from 
black holes and neutron stars, separately.

\subsection{Black Holes}
Since a single black hole cannot shine without environmental gas,
we focus our discussion on the cases of binaries containing black holes,
called black-hole binaries (BHBs).
Spectral properties of BHBs have been investigated rather extensively 
recently and a variety of spectral states have been recognized.
The most well-known spectral states are the so-called high-soft state
and the low-hard state.\cite{rf:tanshi}\cite{rf:mccrem}

In the high-soft state, the disk spectra are blackbody and can well
be represented by the standard disk model\cite{rf:shasun}.
For a given mass $M$ of a black hole and rate $\dot M$
of mass accretion onto a compact star, the blackbody temperature
($T_{\rm disk}$) of an accretion disk at a distance $r$ from the
black hole is given by the following relation:
\begin{equation}
\label{Tdisk}
    \sigma T_{\rm disk}^4 = \frac{3}{8\pi}\frac{GM {\dot M}}{r^3}
                 \left(1-\sqrt{r_* \over r}\right),
\end{equation}
where $r_*$ is the radius of the inner edge of the disk.
It is easy to show that the disk temperature reaches its maximum 
at $r=(7/6)^2 r_*$ and that the maximum temperature is
about $\sim 10^7$ K, as expected,
The standard-type disks are optically thick, thus emitting blackbody
radiation.  Thus, the soft-state spectra are a sum of the blackbody radiation
spectra with multiple temperatures, since the disk temperature is a function
of radius.
The flux will show an exponential roll-over at a frequency,
corresponding to the maximum disk temperature, $\sim 10^7$K.

When the luminosity is less, say, $L/L_{\rm E} < 0.03$, 
where $L_{\rm E}$ 
[$=1.3 \times 10^{38}(M/M_\odot)$erg s$^{-1}$] is the Eddington luminosity,
the spectra become significantly harder (with strong hard X-rays).  
This state is called a low-hard state.
Typically, the disk spectra in the low-hard state 
are modeled by a cut-off power-law;
$f_\nu \propto \nu^{-p} \exp(-h\nu/E_{\rm cut})$,
where $p$ ($\sim 1.7$ typically) is a constant called the spectral index
and $E_{\rm cut}$ ($\sim 100$ keV, which corresponds to a temperature
of $10^9$ K) is the cut-off energy.  
The most promising model explaining the physical situation of accretion 
disks in the low-hard state is the so-called RIAF (radiatively inefficient
accretion flow) model.\cite{rf:naretal}\cite{rf:katoetal}
According to the RIAF model, the disk (or flow)
is not dense enough to emit substantial radiation.
Then the disk gas does not cool efficiently, and 
gets hotter and hotter as one approaches
the central black hole.  As a result,
the disk expands vertically and becomes low-density.

In addition to the two usual states mentioned above, 
two more distinct spectral states are known.
In the very  high state, both
blackbody and power-law components are clearly present with rough equality.
In the slim-disk state\cite{rf:abram}
which appears at even higher luminosities comparable to
or even exceeding the Eddington luminosity\footnote{The maximum
possible luminosity for a spherically accreting system.  No accretion
is possible above this limit, since then radiation pressure force
overcomes gravitational force.}, again disk spectra are of blackbody type
but with flatter temperature profiles

To summarize, we can roughly specify the black-hole mass (or $L_{\rm E}$)
 through the X-ray spectral shapes of the BHBs.
X-ray observations with good sensitivity are
of great importance for this reason.

\subsection{Neutron Stars}
We can classify neutron stars into two categories according to their
energy sources:
rotation powered neutron stars and accretion powered neutron stars.
If neutron stars have strong magnetic fields and rapid rotations
they can emit periodic radio emission by magnetic dipole radiation.
These are observed as pulsars.
The typical magnetic field strengths of pulsars are
$B \sim 10^{12}$ G.  However, recently pulsars have been found
those with extremely large magnetic field strengths,
$B\sim 10^{15}$ G; these are called magnetars.
They are bright in X-rays and occasionally
emit bursts of $\sim 10^{41}$ erg s$^{-1}$
or more in X-ray and gamma-ray ranges. 

The other category, accretion-powered neutron stars
are found in X-ray binaries.  Typically, 
there are two spectral components, both of blackbody type,
observed in X-ray ranges: 
emission from an accretion disk and from the surface of a neutron star.
(Note that a companion star is normally bright 
in the optical and infrared bands.) 
The former shows a spectrum of around 1 keV 
(corresponding to a temperature of $\sim 10^7$ K) 
and the latter of $\sim 2$ keV.  

The virial theorem tells us that half of the gravitational energy 
that is released from gas reaching the neutron-star surface 
at $r_*$ goes to kinematic (rotation) energy,
while the remaining half goes to radiation from the disk surface.
Therefore, the temperature differences of both components
can be simply understood in terms of the
different size of the emitting surface; that is,
the neutron star surface is hotter than the hottest part of the disk
because of its smaller surface area.

\section{Cosmic Laboratories}

The study of dense stellar systems can be taken a long way with the
classical gravitational N-body model.  True, stars are not
point-masses, but this idealization remains the basis of much of our
understanding.  This is the province of
{\sl stellar dynamics}.\cite{rf:3}\cite{rf:bt}\cite{rf:spitz}
In particular we are in a regime somewhere between $N=2$ and
$N=\infty$: the first case can be solved exactly, while in the second
limit a stellar system behaves in most respects like a continuum.
Outside astrophysics, the closest analogy to the regime of interest in
this review is a collisional unmagnetized plasma, with the added
complication of spatial inhomogeneity.

What makes the problem tractable is the empirical fact that a stellar
system of negative total energy is found to settle down reasonably
quickly to a quasi-equilibrium structure.  We shall consider the
simplest, spherically-symmetric case.  An application of the virial
theorem then shows that, in an average sense,
$$
2T + W = 0,
$$
where $T,W$ are, respectively, the kinetic and gravitational potential
energies of the system.  In this situation we can define a length
scale for the system, $R$, called the ``virial radius'', by the equation
$$
W = -GM^2/(2R),
$$
where $M$ is the mass of the system, and $G$ is the universal constant
of gravitation.  $R$ is of the order of a parsec for star clusters.
This relation holds for all isolated self-gravitating systems, stars
as well as star clusters and galaxies.

In this ``virial equilibrium'', the time scale on which stars orbit
within the cluster is of order $2R/v$, where $v$ is the root mean
square speed of the stars (which is expressible in terms of $M$ and
$T$.)  This time scale is called the ``crossing time'', denoted
$t_cr$, and is of the
order of a million years for star clusters.

Virial equilibrium is a dynamical equilibrium, but not (in an
appropriate sense) a {\it thermal} one.  Just like atoms in a gas, stars
in a cluster can exchange energy, though they do so by gravitational
encounters only (at least, in the idealization of point masses).  The
time scale on which they do so is called the {\sl relaxation time},
and it generally much longer than the
crossing time.  In fact in virial equilibrium it is given to order of
magnitude by
$$
t_r \simeq 0.1 N t_{cr}/\ln N,
$$
where $N$ is the number of stars.  

On this time scale, one would naively imagine that a stellar system
can reach thermal equilibrium, the mean square velocities of the stars
being the same everywhere, and inversely proportional to the stellar
mass.  Actually the thermodynamic study of this problem is a
fascinating one\cite{rf:pad}, though it is complicated by
the fact that stellar systems are ``open''.  It is true that
gravitational encounters have the tendency to promote thermal
equilibrium, but the fact that these systems are self-gravitating
makes the results counter-intuitive.\cite{rf:hach}\cite{rf:makhut}
The fact that systems are open implies that stars escape, and so can
never fully populate a Maxwellian velocity distribution.  The fact
that they are self-gravitating implies that, in the exchange of
thermal energy, they behave in some respects as if they had negative
specific heat.  Therefore (though the argument is a little
complicated\cite{rf:lbw}), stellar systems tend to form a hot, dense
core, in a process referred to as ``core collapse''.  Its time scale
is typically $15t_r$ if all stars have the same mass.

This is an unrealistic idealization, however, and we consider now the
effects of the presence of particles with a variety of masses, which
exhibit a effect of the tendency to equipartition of velocities: the
heavier stars tend to lose kinetic energy in encounters, and sink into
the core\cite{rf:spi}, and they do so on a time scale
of order $t_r<m>/m_{max}$, where $<m>$ is the mean mass, and $m_{max}$
is the mass of the heavier stars.

In real stellar systems, many stars are double (like a gas consisting
of a mixture of single atoms and diatomic molecules).  Three- and
four-body interactions between these components introduce a new
feature:  a tendency (in interactions involving close pairs) for the
kinetic energy of the products of an encounter to exceed that of the
original participants.  These interactions behave like an exothermic
chemical reaction, heating up the gas of stars.  This brings the
process of core collapse to a halt (in systems with stars of a single
mass), and has a similar effect on the segregation of heavy stars
toward the center of the system.  This puts the system at last onto a
kind of thermal equilibrium, rather like that of a star, in which the
central production of energy supports and powers the entire system,
whose stars gradually evaporate, like enormous photons, into the
environment of the stellar system.

To add realism to this fascinating but still idealized picture, we
have to add the details of the way in which individual stars evolve
internally while these dynamical processes are taking place.  We do so
in the context of astrophysical systems where this interplay is
occurring before our eyes.

In the vicinity of the Milky Way galaxy are several places where the
interaction between stellar and dynamical evolution has made its
mark.  Best studied of all are the old globular star clusters, because
they are the closest and least obscured.  Somewhat further away is the
Galactic Center, though detailed observations became possible only
recently because it is heavily obscured by dust and gas.  Still
further off are the young dense star clusters where stars have
recently formed.  Examples also occur in the Milky Way but again
behind heavy layers of obscuration.

\subsection{The Galactic Center}

\begin{figure}
\centering
\includegraphics[angle=0,width=\textwidth]{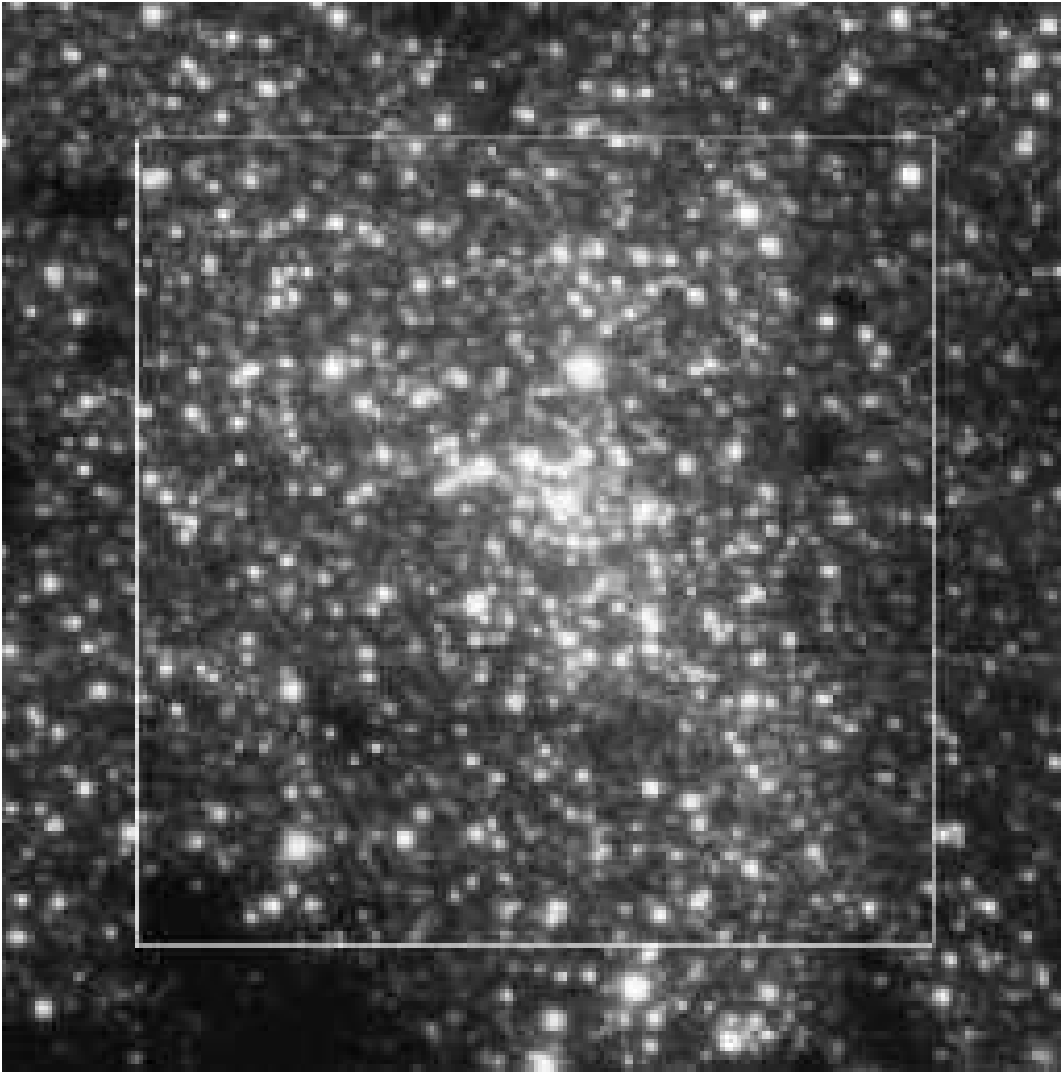}
\caption{The star cluster at the Galactic Center\cite{rf:b}.
  The width of the outer frame is about 3 pc.}
\end{figure}

The center of our galaxy lies at a distance of about 8 kiloparsecs in
the southern constellation of Sagittarius.  It is marked by a bright
radio and X-ray source, Sagittarius A*, which is now widely considered
to be a black hole with a mass of approximately $3\times10^6M_\odot$.
It is surrounded by a ``nuclear star cluster'' in which the stellar
density is well approximated by a broken power law with a break radius
at about 0.22 parsec from the black hole\cite{rf:a}.  At this radius
the stellar density is about 3$\times10^6M_\odot$/pc$^3$, and an
additional comparable mass density is thought to be present in the
form of stellar remnants (black holes, neutron stars, white dwarfs).

At the break radius, the time of relaxation (which is the time scale
on which two-body gravitational encounters are effective) is already
less than 1Gyr\cite{rf:c}, or about one tenth the age of the
universe.  Physical collisions become important on this time scale
only much closer to the black hole, at a distance of order
0.02pc\cite{rf:d}.  The short two-body relaxation time suggests that
heavier stars near the galactic center should have sunk toward the
center relative to lighter stars, by a process called ``mass
segregation''.  (If the velocity dispersion of stars is independent of
mass, an encounter between stars of different mass tends to leave the
more massive star with a smaller velocity, which causes it to move
onto a smaller orbit.)  Indeed it is found that stars very close to
the black hole are very massive, but they are also very {\sl young}.
Therefore they have not existed for long enough,  compared with the
relaxation time, to exhibit mass segregation.  They must have appeared
close to the black hole by some other processes.  

They could have formed {\sl in situ}.  But the enormous tidal
gravitational field in the vicinity of the black hole should be an
insurmountable obstacle to star formation.  Perhaps they arose by
collisions between more normal stars, which coalesce, resulting in a
more massive, young-looking star?  As we shall see, however, at
the velocity dispersion found near the galactic center collisions
destroy stars and do not make heavy stars.  Another possibility is
that the stars were born further away from the galactic center and
rapidly migrated there.  For example, a binary star may pass close to
the black hole, which could disrupt the binary, and leave one star orbiting
in the vicinity of the black hole.  Indeed, at larger distances from
the galactic center, where star formation presumably {\sl is}
possible, there are a few massive young star clusters.  As these
evolve and disrupt they tend to move toward the galactic center (by a
process akin to mass segregation), and could deliver young stars to
its vicinity, though it is hard to see how this can be done quickly
enough.

There is another possible explanation for the origin of S stars which
simultaneously accounts for another puzzle about the stellar
populations there: the observed paucity of red giants (relative to
other types of stars) within about 0.2 pc of the galactic center.
While collisions between binary stars and red giants may play a role
\cite{rf:dbbs}, another possibility is that the envelopes of red
giants are stripped off in encounters with the black hole.  If so,
something resembling an S star might result \cite{rf:dk}.

There is another effect of all this dynamics near the black hole.
Whether or not disrupted binaries can account for the S-stars, stars
emerging from disruptive encounters can leave the galactic center with
high speeds, much higher than the speeds of any other single stars in
the Milky Way.  Indeed in recent years several such high-speed stars
have been discovered, though it is not clear that all of them can have
emerged from the galactic center.

The Galactic Center can be studied in considerable detail because it
is so close.  Most galaxies are thought to harbour black holes, but
their effects on stellar populations are much harder to detect.  It
has been proposed, however, that a recently discovered population of
hot stars in the vicinity of the center of M31 (the Andromeda Galaxy)
may share their origin with the S stars in the Galactic Center,
controversial though that still is.  

The Andromeda galaxy is more distant than the Galactic Center by a
factor of about a hundred.  Still further away are many examples of
galaxies also thought with high probability to harbour black holes.
It is impossible to detect the effects of these black holes on the
stellar populations, except for the important dynamical effects on the
space distribution and kinematic properties of the stars.  But these
black holes are an important indicator of the manner in which galaxies
are formed, for it is found that there is a strong correlation between
the inferred mass of the black hole and that of the galaxy itself (or,
to be more precise, a particular component of it - the so-called
``bulge'').  The most massive galaxies contain the most massive black
holes, with black hole masses up to around $10^9M_\odot$.

\subsection{Globular Star Clusters}

The study of stellar populations in dense galactic nuclei is so hard
because they are so distant or so highly obscured, or both.  Globular
star clusters are dense stellar systems which suffer from neither
problem.  They are smaller, mostly with less than a million stars
each, but the closest one (the star cluster M4) is only about one
quarter of the distance to the Galactic Center.  These are the objects
of choice for studying the effects of stellar density (number of stars
per unit volume) on stellar populations. 
In the following, we discuss the exotic objects found in globular
clusters, such as blue stragglers (4.2.1), X-ray binaries (4.2.2),
unusual stellar populations (4,2.3) and black holes (4.2.4).

\begin{figure}
\centering
\includegraphics[angle=0,width=\textwidth]{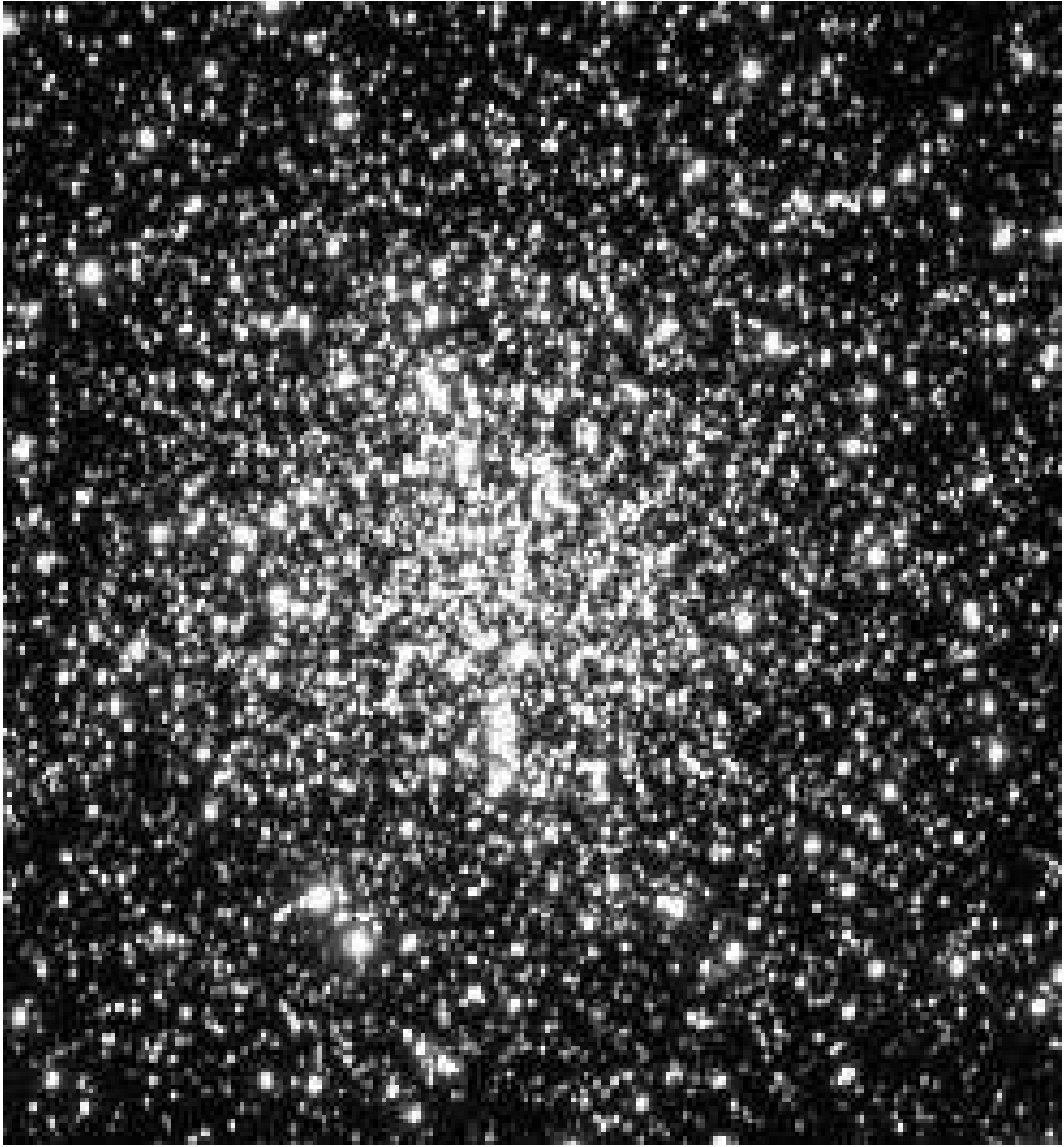}
\caption{The nearest globular star cluster, M4, which contains roughly
$10^5$ stars.  The width of the frame is about 4 pc.  (Credit NOAO/AURA/NSF)}
\end{figure}

\subsubsection{Blue stragglers}\label{sec:bs}

The first indications came over 50 years ago from studies of the
colors and magnitudes of stars in individual clusters.\cite{rf:sand}
Each cluster appeared
to consist of stars of a unique age (comparable with the age of the
Milky Way itself), except for a small number of ``blue stragglers''.
As their name suggests, they are hotter stars than the bulk of the
stars in the cluster, and they {\sl look} relatively youthful, to
judge by their luminosity and color.  In this respect they somewhat
resemble the S stars near the Galactic Center, and they pose the same
problems.  Though they might have formed more recently than the other
stars, there is no sign in any globular cluster of the gas clouds
from which young stars can be born, and there are good reasons why
such gas cannot exist in sufficient quantities.  Unlike the S stars,
however, there are no essential difficulties in supposing that at
least some of the blue stragglers are the result of collisions; the
velocity dispersion in a globular cluster is much smaller than in a
galactic nucleus, and colliding stars should coalesce with little
loss of mass.\cite{rf:bail}
The main difficulty is in establishing that collisions
are common enough and the result will look like a blue straggler.

Estimates show that collisions in star clusters {\sl are} rather
common.  In the globular cluster 47 Tucanae, for instance, perhaps as
many as 1000 new systems (about 0.1\% of all stars in the cluster) are
produced by collisions between single stars in its
lifetime\cite{rf:e}.  Not all of these are blue stragglers, but the
numbers are enhanced by the behavior of {\sl binary stars}.  Once
thought to be rare in globular clusters, it is now considered that
their abundance is not much less than in the vicinity of the sun
(i.e. over 50\%).  When one of the stars in a binary evolves, after
consuming its central supply of hydrogen, it expands, and the two
stars may well coalesce into a single star, which under suitable
circumstances might be interpreted as a blue straggler.

Much effort is now being expended in conducting complete censuses of
blue stragglers in globular clusters, to try to disentangle these two
effects.  Note that both are influenced by the dense stellar
environment.  For stellar collisions this is obvious, but also a dense
environment may destroy the binaries which otherwise, in the course of
time, would form blue stragglers.  This explains why the number of
blue stragglers in a cluster is found to be in anticorrelation with
the number of collisions\cite{rf:f}.

One of the most intriguing problems posed by the collision hypothesis
is how the collision remnant will evolve.  At birth it is a highly
unusual object, with an anomalous mixture of elements (depending on
the state of evolution of the colliding stars), a high rotation rate
(because of their orbital angular momentum), and high internal energy
(from their original relative kinetic energy).  This is one of the
main motivations for trying to build a model of a star cluster which
places dynamics, stellar evolution and collision hydrodynamics on an
equal footing.  All three ingredients have a strong role to play in
determining the outcome of a collision.

\subsubsection{X-ray sources}

Historically the next indication that the stellar environment in a
star cluster influences the stellar population there came from
high-energy astrophysics.  In the early days of X-ray astronomy it was
found that sources were correlated with the positions of globular star
clusters\cite{rf:g}, a star in a cluster being about 100 times more likely to be
an X-ray source than stars outside clusters.  After two or three wrong
turns, it was realized that these sources were binary stars containing
a neutron star, the X-rays resulting from accretion of material from
the outer layers of the binary companion.  The reason why such objects
are rare outside clusters is that a star in a binary is liable to
destroy the binary when the star turns into a neutron star\cite{rf:h} (at the
end-point of its stellar evolution).  In a cluster, however, it is
relatively easy for a single neutron star to interact gravitationally
with a binary (in a three-body interaction) in such a way that the
neutron star displaces one of the original components of the binary\cite{rf:i}.
Such events would be extremely rare outside clusters.  This remarkable
overproduction of X-ray binaries in clusters has the effect that, when
one surveys nearby external galaxies, a large fraction of the bright X-ray
sources one finds are associated with rich star clusters\cite{rf:j}.  

There is another, weaker, class of X-ray sources (called {\sl
cataclysmic variables}) in which the role of the neutron star is taken
by a white dwarf.  Gravitational encounters are again invoked to
explain their numbers, and quite recently this explanation was rather
directly confirmed by the discovery of a strong correlation between
the numbers of discrete X-ray sources in a cluster and the number of
gravitational encounters\cite{rf:k}.

\subsubsection{Anomalous populations}

Another class of ``star'' found in extraordinary abundance in globular
clusters are pulsars (neutron stars), especially those of the
``millisecond'' variety (i.e. the most rapidly rotating, as the name
suggests).\cite{rf:l}  Again the process of capture of a neutron star
by a normal binary is the key to understanding why they are found
predominantly in places where the stellar density is high \cite{rf:m}.

While various details of the populations of blue stragglers, X-ray
sources and millisecond pulsars remain to be clarified, the broad
outlines are clear.  But there are other features of stellar
populations in star clusters which are not understood at all.  Some
time ago it was found that there are color gradients in some
clusters, the center being bluer than the periphery, and it was
established that this was associated with a depletion of red giant
stars near the center and with the density of the central ``core'' of
the cluster\cite{rf:n}.  Perhaps this is due
to the fact that encounters between
stars are sufficiently frequent that the weakly bound envelopes of red
giants are stripped,\cite{rf:o}\cite{rf:p}
but the argument has not been made quantitative
and the problem itself has fallen into neglect.

Another feature which defies explanation is the behavior of the
``horizontal branch''.  This is a distinctive, bright and hot
sub-population of evolved stars, which is found in all globular
clusters.  In some, however, this group of stars includes a number of
faint and hot members called ``faint blue horizontal branch stars'' \cite{rf:q}.
Again they appear to correlate with the density of the central part of
the cluster.  Even more puzzling, however, is the recent discovery
that the stars at the red extreme of this sequence 
are rotating very fast\cite{rf:r}.  The problem of explaining this peculiar
population is wide open, but one possible explanation of fast rotation
is coalescence following a collision (see the discussion of
collisions in Sec.\ref{sec:bs}).  

\subsubsection{Black holes}

Finally we turn to the role of black holes.  Stellar-mass black holes
form in globular clusters as a natural end-product of the evolution of
massive stars.  They probably have a role to play in shaping the core
of a star cluster (which may be defined as the region near the center,
where the stellar density is within a factor two, in projection, of
the central value).\cite{rf:s}\cite{rf:t}  But here we are thinking of {\sl
intermediate-mass} black holes, a putative class of black holes with
masses between those of stellar remnants and those of the supermassive
black holes in galactic nuclei; that is, black holes with a mass of
order 1000$M_\odot$.

One argument for supposing that such objects might exist in globular
clusters is the known correlation between black hole mass and galaxy
bulge mass in galaxies; if this correlation is extended down to the
mass of a globular star cluster, the inferred mass of a black hole
would be of order 1000$M_\odot$\cite{rf:u}.  It is not clear that globular
clusters are a low-mass extension of galaxies, but a few of the most
massive clusters (such as $\omega$ Centauri) are often interpreted as
being the nuclear remnants of small galaxies.\cite{rf:v}\cite{rf:w}

More direct evidence for intermediate-mass black holes in a few star
clusters has come from measurements of the surface brightness and
velocity dispersion of the stars, in much the same way that galactic
supermassive black holes are identified.  But the interpretation is
complicated by the effects of mass segregation, which tend to modify
these profiles in much the same way as a black hole does, and so the
existence of an intermediate mass black hole is not uncontroversial in
any cluster.\cite{rf:x}\cite{rf:y}\cite{rf:z}

A third form of evidence comes from rich star clusters of a very
different kind.  These are the young dense star clusters (dubbed
yodecs) which are found in abundance in some kinds of galaxies
(interacting galaxies, starburst galaxies) where the formation of
stars is taking place at much higher rates than in the
Milky Way(e.g.\cite{rf:aa}).  A
few of these are found to be X-ray sources with a luminosity which
suggests that the accreting object is a black hole (rather than a
neutron star or white dwarf)\cite{rf:bb}.

The next question to be addressed is how such black holes can arise.  
Computer Modeling of yodecs has delineated the conditions under which
this appears to be possible.\cite{rf:cc}\cite{rf:dd}
In its simplest terms, the condition is
that a star must be able to collide and coalesce with other stars on
a time scale no longer than the time of evolution of a massive star
(i.e. a few million years).  Then a very massive star may build up by
a process of runaway coalescence.  This condition on the collision
rate requires a very high space density of the stars, but it appears
that it can be satisfied in the densest yodecs, especially if the
dynamical segregation of high mass stars is taken into account.  The
most uncertain step in this scenario for the formation of intermediate
mass black holes, however, is what happens next:  does an extremely
massive star give rise to an intermediate-mass black
hole?\cite{rf:ee}\cite{rf:ff}

This review of dense stellar systems has now come full circle, because
one object which may harbour an intermediate mass black hole is group
of stars close to the Galactic Center.  These stars may be the most
massive members of a cluster, and their velocity dispersion implies
that, if they are indeed gravitationally bound, the presence of an
intermediate mass black hole is indicated.\cite{rf:gg}\cite{rf:hh}
Certainly, elsewhere in
the vicinity of the Galactic Center there are a number of bona fide
star clusters\cite{rf:ii}.  During their short lives they spiral toward the
Galactic Center itself.  If some of these also contain
intermediate-mass black holes, they may contribute to the build-up of
the central black hole itself.\cite{rf:jj}\cite{rf:kk}

\subsection{Stellar Collisions}

The foregoing review of cluster dynamics focused on issues involving
three processes: stellar evolution, stellar dynamics, and the
non-gravitational interactions between stars.  Now that the study of
such interactions has been motivated, we shall describe in a little
more depth just what is involved in modeling this problem.  Of the
three processes, it is the least well developed.

Long ago Spitzer and Saslaw \cite{rf:ll} proposed a model which works rather when
the relative velocity of the two interacting stars is high (relative
to the escape speed from their surface).  Roughly speaking, the
parts of the stars which overlap (when viewed along the direction of
their relative motion) is assumed to be lost if its total energy
(calculated using momentum and energy conservation) is positive.  The
condition of high relative velocity is reasonably well satisfied in
galactic nuclei.

For low-velocity encounters, such as generally occur in globular star
clusters, the best developed analytical approach is one due to
Lombardi and his colleagues\cite{rf:mm}.  First the entropy profiles of the two stars are calculated,
and then their union is sorted by specific entropy to construct the
radial profile of the coalesced star.

Beyond these two approaches, hydrodynamical simulations are the method
of choice, and again there are two approaches.  One is through the use
of grid-based hydrodynamic codes (as used in this problem, for
example, by Ruffert\cite{rf:nn}), and the other, now perhaps more comprehensive,
is Smoothed Particle Hydrodynamics.  The latter approach has been
developed over many years in the astrophysics community and several
sophisticated aspects of its behavior are now well established in
various contexts.\cite{rf:oo}\cite{rf:pp}

The result of an encounter depends in an essential way on several
factors: the masses of the two stars, their stellar type (red giants
and neutron stars behave in very different ways, for example), their
relative speed, and their distance of closest approach.  Extensive
libraries of collision data can now be accessed\cite{rf:qq}, but
there are other important parameters which render the tabulation of
collision outcomes increasingly difficult, e.g. stellar rotation.  For
this reason, there is a real need to incorporate live SPH collision
simulations into a comprehensive program for simulating all essential
aspects of dense stellar systems (Sec.5).

Collisions between stars are rare events, even in dense stellar systems.
Recently, however, an unusual observation was made that could possibly
be caused by a stellar collision, in the galaxy M85, in the Virgo
cluster\cite{rf:pr}.

\section{Virtual Laboratories}

Astrophysics is the only field within physics that has no laboratory
component.  Within the immediate neighborhood of the Earth, in our own
solar system, we have access to meteorites that reach the Earth and
samples that have been returned from the Moon or that have been
analyzed {\it in situ} by robotic explorers on planets or moons in
our solar system.  But anything outside our own planetary system is
completely outside our reach.

Until half a century ago, astronomy thus had the strange distinction
of being at the same time the oldest modern science, giving rise to
classical mechanics, but also the only modern science without a lab.
Happily, this changed with the advent of electronic computers, which
have provided astrophysicists with a virtual laboratory in which to
conduct experiments.

\subsection{Early History}

The first experiments conducted in these laboratories focused on the
evolution of single stars.  Models of stars such as our own Sun were
constructed, and they were evolved during the billions of years of
their total life time.  Within a decade, the field of computational
stellar evolution matured enough to see the publication of several
text books.

With a somewhat later start, in the early sixties, simulations of the
gravitational interactions between stars in star clusters took off.
This field of stellar dynamics is similar in many respect to that of
molecular dynamics, in the classical approximation.  In both cases,
interesting results can be obtained in the point-particle approximation,
and the main difference is the use of an attractive inverse square
force law to model gravity, versus more complex laws such as those
based on a Lennard-Jones potential in the case of molecules.

It is an interesting question why these two fields, stellar evolution
and stellar dynamics, remained relatively separate for several decades,
Only in the last ten years have stellar dynamics simulations included
detailed recipes for the evolution of single stars and double stars,
and a real coupling of `live' stellar evolution codes and stellar
dynamics codes has only begun last year, with the work of Church\cite{rf:1}.

A third area that is relevant is that of stellar hydrodynamics,
describing what happens when two stars physically collide.  The very
first calculations in this field were also done in the sixties, but
again, stellar hydrodynamics did not become integrated with stellar
evolution and stellar dynamics for a long time.  In fact, we are still
waiting for the first such combined simulations.

\subsection{Multi-Scale Challenges}

In the introduction, we have already mentioned the vastly different
spacetime scales on which we have to model the evolution of a star
clusters.  Spatial scales range over 15 orders of magnitude, from
kilometers to parsecs, and temporal scales range over 21 orders of
magnitude from fractions of milliseconds to many billions of years.
As a result, it is impossible to follow the evolution of a star
cluster using standard text book numerical integration schemes,
even if we were to approximate stars as point particles.

To see this, let us make a simple estimate of the computational needs
for a straightforward stellar dynamics simulation.  In a globular
cluster with a million stars, there are $10^{12}$ pair-wise
gravitational forces that we have to consider.  If we were to evaluate
each pair-wise force, on the shortest time scale that configurations
can change, we would need to repeat that exercise $10^{21}$ times.
This implies that we need to calculate $10^{33}$ pair-wise force
calculations.  A typical calculation involves a few dozen floating
point calculations, so the total cost would exceed $10^{34}$ floating
point calculations.  Even with a future supercomputer speed of
1 Petaflops, or $10^{15}$ floating point calculations per second,
a simulation would take $10^{19}$ seconds, or more than $10^{12}$
years, a hundred times longer than the current age of the universe.

Of course, it would be an enormous waste of time to model a star
cluster in such a way, using constant time steps.  A much better
approach would be to use adaptive time steps.  At each moment, we can
determine which stars are involved in a relatively close encounter,
and we can then enforce a system timestep of the appropriate size.
Sometimes such a timestep will be a fraction of a millisecond, but at
other times the closest stellar passage may take place on time scales
of seconds or minutes, thereby speeding up the calculations by several
orders of magnitudes.  However, this in itself will not beat the
factor of $10^{12}$ that separates us from a brute-force calculation
and the requirement to finish a simulation in at most a year.

Unfortunately, most standard numerical methods textbooks do not go
further than recommending adaptive time steps.  In order to make
stellar dynamics calculations of star cluster feasible at all,
astrophysicists had to invent completely new integration schemes,
together with all kinds of other specialized algorithmic tricks.

A very important step is the switch from shared adaptive time steps to
individual time steps.  When some stars undergo a particularly close
encounter, there really is no reason to slow all other stars down to the
same short time steps as is needed to resolve that encounter.  Stars
that are relatively more isolated can be allowed to take much longer
integration time steps.  The trick here is to use an extrapolation
method to predict the position of those slow stars, to allow us to
calculate their forces on those stars that move faster and in more
irregular ways.

Another equally important step is to treat the evolution of tight
clumps of small numbers of stars separately.  A tight double star,
for example, can be integrated analytically when the stars are so
close together that the perturbations of all other stars can be
safely neglected.  And even when two stars are not that close, the
perturbation of other stars can be taken into account in approximate
ways, saving orders of magnitude in computer time.  Similarly, triple
stars that are almost isolated can be effectively frozen or put in a
form of quarantine until neighboring stars come sufficiently close to
play an important role.

For an overview of various classes of algorithms, see Aarseth\cite{rf:4}.
For a detailed analysis of the computational costs of the main algorithms,
see Makino \& Hut.\cite{rf:011}\cite{rf:012}

\subsection{Multi-Physics Challenges}

\subsubsection{The Basic Picture}

It is easy to sketch a picture of how you could model the structure
and evolution of a star cluster, while taking into account the
structure and evolution of the individual stars that constitute the
cluster.  Since most stars are either single, or part of a binary in
which the stars are not in very close proximity, those stars can be
modeled as point particles, to a high degree of accuracy.  For each
such star, we can run a stellar evolution code to keep track of their
internal properties, as well as their radius, but with respect to the
dynamical interactions with other stars, their non-zero size does not
need to be taken into account.

When two or more stars come close to each other, however, the point
mass approximation breaks down.  In such a case we can take the mass
points, representing the stars involved in such an encounter, and
replace them by hydrodynamical models.  For example, we can use the
Smooth Particle Hydrodynamics (SPH) approximation, which is natural
since it is also particle based, but we could use other approaches as
well.\cite{rf:10}  We can then model the subset of hydrodynamically
realistic stars during the short time of their encounter, under the
perturbing influence of nearby stars that are still treated as mass
points.

On a time scale short with respect to the evolution of the whole star
cluster, typically days or weeks, the two or more stars may merge, or
separate again, or perhaps settle into a contact binary configuration.
In any of these cases, we can wait for dynamic equilibrium to be
restored, and then we can replace the stars again by point masses, as
far as the subsequent stellar dynamical evolution is concerned, with
a special treatment for the higher-order multipoles of the binaries,
where needed.

For the stars to regain their thermal equilibrium will require vastly
longer than the days or weeks needed for dynamical equilibrium to be
restored.  Here we are talking about millions or perhaps hundreds of
millions of years, depending on the mass of the stars.  During this
time we will have to run an active stellar evolution code to model the
evolution of such an unusual star (for other, normal, stars we have
the luxury of using table look-up methods, rather than evolving the
star using a live stellar evolution code\cite{rf:10}).

\subsubsection{Steps Toward Implementations}

The picture sketched above may look simple in principle, but writing a
computer code that can automatically take care of all the conversions
and interactions is a challenging task.  The logistics for treating
encounters between single stars and/or binaries is already quite
complicated.  What is worse, there is a small chance that additional
stars or binaries will show up, after the initial interactions have
started.  For example, when two binaries encounter each other, a third
binary may approach the system, leading to a strongly interacting
six-body system.  To treat all possible types of outcome correctly is not
an easy programming task.

An important step in the direction of constructing such a code was
taken with the release of the Kira code, at the core of the Starlab
environment.\cite{rf:zzy}\cite{rf:zzz}  In this case, stellar
evolution was modeled through the use of recipes and fits to stellar
evolution tracks.  In the near future, actual stellar evolution codes
will be connected to stellar dynamics and hydrodynamics codes.  A step
in that direction is currently being taken by the international MUSE
collaboration\cite{rf:zzx}, aimed at connecting existing codes in all
three fields, using Python as a `glue' language.  The MUSE project is
one of the main activities under the MODEST umbrella, short for
MOdeling DEnse STellar systems\cite{rf:9}.  A complimentary approach
is made in the ACS project, short for the Art of Computational
Science\cite{rf:zzv}.  There the emphasize is on writing extremely
modular codes that have many levels of hooks for connecting with other
modules that model different types of physics.

\subsection{GRAPE}

As mentioned in section 2.4, GRAPEs have been extensively used for the
simulations of dense stellar systems. The first GRAPE hardware used
for such simulations was GRAPE-2\cite{rf:zza}, with a peak speed
of 40 Mflops.  In 1995, the GRAPE-4 system, with a peak speed of 1.08
Tflops, was completed.  It used 36 processor boards, each with 48
special processor chips.  One processor chip delivered a speed of 640
Mflops.  Thus, a single processor board had a speed of about 30
Gflops.  In 2002 the GRAPE-6, with a peak speed of 64 Tflops, was
completed.  Its processor chip has a speed of 31 Gflops, and the peak
speed of a single board with 32 processor chips is 1 Tflops.  A 4-chip
version of the GRAPE-6 was also developed.  Many copies of GRAPE-6 boards
were made, and they are used by many researchers, around the world.

Currently, the GRAPE-DR system is under development.\cite{rf:zzs}\cite{rf:zzt}
Unlike previous versions of GRAPE hardware, which have processors
specialized for the calculation of pairwise gravitational interactions
between particles, GRAPE-DR integrates a number of very simple but
programmable processors, and can be used for a much wider range of
applications. The name GRAPE is for Gravity Pipe, but GRAPE-DR means
Greatly Reduced Array of Processor Elements with Data Reduction.
An obvious application for the GRAPE-DR processor are SPH calculations.
A single GRAPE-DR processor chip delivers 512 Gflops, and a small 4-chip
card  will deliver 2 Tflops, or twice the speed of the large, 32-chip
GRAPE-6 board.

\section*{Acknowledgments}
DCH was a recipient of a JSPS Visiting Fellowship (No. 07031) while working on
this review, and thanks his host, SM, for kind hospitality during that period.
P.H. thanks Profs. Masao Ninomiya and Shin Mineshige for inviting him
to visit the Yukawa Institute for Theoretical Physics, at Kyoto University.
This work was supported in part by the Grant-in-Aid for the 21st Century COE 
``Center for Diversity and Universality in Physics'' from the Ministry of
Education, Culture, Sports, Science and Technology (MEXT) of Japan (S.M.).

\def\mnras{MNRAS}
\def\apj{ApJ}
\def\aj{AJ}
\def\araa{Ann. Rev. Astron. Astroph.}
\def\aap{A\&A}
\def\apjl{ApJL}

\end{document}